\documentclass[10pt]{article}
\usepackage{bbm}
\usepackage[final]{graphics}
\usepackage{amsmath}
\usepackage{amsfonts,amsbsy}
\usepackage{amssymb}
\usepackage{bm}
\usepackage{color}

\def\empile#1\over#2{\mathrel{\mathop{\kern 0pt#1}\limits_{#2}}}
\def\bs{\boldsymbol}
\def\wt#1{\widetilde{#1}}

\def\TODO#1{}

\def\p{{\boldsymbol p}}

\newcommand{\slL}{\raise.15ex\hbox{$/$}\kern-.53em\hbox{$L$}}
\newcommand{\slP}{\raise.15ex\hbox{$/$}\kern-.53em\hbox{$P$}}
\newcommand{\slD}{\raise.15ex\hbox{$/$}\kern-.67em\hbox{$D$}}
\newcommand{\slp}{\raise.1ex\hbox{$/$}\kern-.63em\hbox{$p$}}
\newcommand{\slq}{\raise.1ex\hbox{$/$}\kern-.53em\hbox{$q$}}
\newcommand{\slv}{\raise.1ex\hbox{$/$}\kern-.63em\hbox{$v$}}
\newcommand{\slR}{\raise.15ex\hbox{$/$}\kern-.53em\hbox{$R$}}
\newcommand{\slQ}{\raise.15ex\hbox{$/$}\kern-.53em\hbox{$Q$}}
\newcommand{\slK}{\raise.15ex\hbox{$/$}\kern-.53em\hbox{$K$}}
\newcommand{\slk}{\raise.15ex\hbox{$/$}\kern-.53em\hbox{$k$}}
\newcommand{\slSigma}{\raise.15ex\hbox{$/$}\kern-.53em\hbox{$\Sigma$}}
\newcommand{\slcalP}{\raise.15ex\hbox{$/$}\kern-.63em\hbox{$\cal P$}}
\newcommand{\slcalA}{\raise.15ex\hbox{$/$}\kern-.63em\hbox{$\cal A$}}
\newcommand{\slA}{\raise.15ex\hbox{$/$}\kern-.73em\hbox{$A$}}
\newcommand{\slbfA}{\raise.15ex\hbox{$/$}\kern-.73em\hbox{${\imb A}$}}
\newcommand{\slpartial}{\raise.15ex\hbox{$/$}\kern-.53em\hbox{$\partial$}}
\newcommand{\sla}{\raise.15ex\hbox{$/$}\kern-.53em\hbox{$a$}}
\newcommand{\slb}{\raise.15ex\hbox{$/$}\kern-.53em\hbox{$b$}}
\newcommand{\slc}{\raise.15ex\hbox{$/$}\kern-.53em\hbox{$c$}}
\newcommand{\slC}{\raise.15ex\hbox{$/$}\kern-.63em\hbox{$C$}}
\newcommand{\sln}{\raise.15ex\hbox{$/$}\kern-.575em\hbox{$n$}}

\newcommand\ontop[2]{\genfrac{}{}{0pt}{}{#1}{#2}}

\begin{document}

\title{\bf From lattice Quantum Electrodynamics\\
 to the distribution of the algebraic areas\\
 enclosed by random walks on ${\mathbbm Z}^2$}
\author{Thomas Epelbaum${}^1$\footnote{tomepel@gmail.com}, Fran\c cois Gelis${}^2$\footnote{francois.gelis@cea.fr (corresponding author)}, Bin Wu${}^2$\footnote{bin.wu@cea.fr}}

\maketitle

\begin{itemize}
\item[{\bf 1}.] McGill University, Department of Physics\\ 3600 University Street, Montreal QC H3A 2T8, Canada
\item[{\bf 2.}] Institut de physique th\'eorique, Universit\'e Paris Saclay\\
 CEA, CNRS, F-91191 Gif-sur-Yvette, France
\end{itemize}

\begin{abstract}
  In the worldline formalism, scalar Quantum Electrodynamics on a
  2-dimensional lattice is related to the areas of closed loops on
  this lattice.  We exploit this relationship in order to determine
  the general structure of the moments of the algebraic areas over the
  set of loops that have fixed number of edges in the two
  directions. We show that these moments are the product of a
  combinatorial factor that counts the number of such loops, by a
  polynomial in the numbers of steps in each direction. Our approach
  leads to an algorithm for obtaining explicit formulas for the
  moments of low order.
\end{abstract}
\begin{flushleft}
MSC~:~05A15\\
Key~words~:~random walks on ${\mathbbm Z}^2$, algebraic areas, lattice QED
\end{flushleft}

\section{Introduction}
In a recent work \cite{EpelbGW2}, we have studied the short distance
behavior of the one-loop expectation value of local operators in
lattice scalar quantum chromodynamics. For this study, we used a
discrete version of the worldline
formalism~\cite{BernK1,BernK2,Stras1,Schub2}, in which propagators are
represented as a sum over all the random walks on the lattice, that
connect the two endpoints of the propagator. In this approach, the
coefficients of the expansion in powers of the lattice spacing are
related to combinatorial properties of these random walks. The leading
coefficient merely counts the random walks of a certain length, but
the following coefficients are related to areas enclosed by the random
walk (for local operators, the two endpoints of the propagator are at
the same point, and therefore its worldline representation is a sum
over closed random walks). In fact, the successive terms of this
expansion can be related to the moments of the distribution of the
areas. In the case of an isotropic lattice (i.e. with identical
lattice spacings in all directions), the results we needed could be
found in ref.~\cite{MingoN1}, in which the authors obtain the general
structure of the moments of the area distribution over the set of
closed random walks of fixed length $2n$. Moreover, this paper
contains explicit formulas for the moments of order 2 and 4, that were
sufficient for our purposes in the isotropic case.

However, the generalization of this expansion to the case of an
anisotropic lattice (i.e. with different lattice spacings in each
direction) requires the moments of the area over the set of closed
random walks that make $n_1$ steps in the direction $+1$ and $n_2$
steps in the direction $+2$ (and also $n_1, n_2$ steps in the
directions $-1,-2$). Some numerical explorations and guesswork
suggested a very simple form for these moments, namely the product of
the number of such closed walks times a polynomial in $n_1,n_2$. Then,
by taking this structure for granted, we obtained the expressions of
these polynomials for the moments $2,4,6,8$ and $10$ by making an
exhaustive list of all the closed random walks up to the length
$2n=20$. The purpose of the present paper is to prove these formulas,
by exploiting the relationship between the distribution of areas and
Quantum Electrodynamics (QED) on a 2-dimensional lattice.

Let us consider an infinite planar square lattice, and denote by
${\bs\Gamma}_{n_1,n_2}$ the set of the closed loops (with a fixed base
point to avoid considering loops that are identical up to a
translation) drawn on the edges of this lattice, that make $2n_1$ hops
in the directions $\pm1$ and $2n_2$ hops in the directions $\pm2$.  We
are interested in the distribution of the algebraic areas enclosed by
the loops in ${\bs\Gamma}_{n_1,n_2}$. In this paper, we prove that
\begin{equation}
\sum_{\gamma\in{\bs\Gamma}_{n_1,n_2}}
\!\!\!\!\!\!
\left({\rm Area}\,(\gamma)\right)^{2l}
=\frac{(2(n_1\!+\!n_2))!}{n_1!^2n_2!^2}\;{\cal P}_{2l}(n_1,n_2)\; ,
\label{eq:moments-0}
\end{equation}
where ${\cal P}_{2l}$ is a symmetric polynomial in $n_1,n_2$ of degree
$2l$. Note that the combinatorial factor
$(2(n_1\!+\!n_2))!/(n_1!^2n_2!^2)$ is nothing but the cardinal of the
set ${\bs\Gamma}_{n_1,n_2}$. Therefore, the polynomial ${\cal
  P}_{2l}(n_1,n_2)$ can be interpreted as the mean value of
$\left({\rm Area}\,(\gamma)\right)^{2l}$ over this set. Our approach
provides an algorithm for calculating these polynomials
explicitly. The first two non-zero moments involve the following
polynomials\footnote{By summing these expressions over all
  $n_1+n_2=n$, we recover the results of ref.~\cite{MingoN1},
\begin{eqnarray*}
&&
\sum_{n_1+n_2=n}\frac{(2(n_1\!+\!n_2))!}{n_1!^2n_2!^2}\;\frac{n_1n_2}{3}
=\left(\ontop{2n}{n}\right)^2\;\frac{n^2(n-1)}{6(2n-1)}\; ,
\nonumber\\
&&
\sum_{n_1+n_2=n}\frac{(2(n_1\!+\!n_2))!}{n_1!^2n_2!^2}\;\frac{n_1 n_2\big({7 n_1n_2}
\!-\!({n_1\!+\!n_2})
\big)}{15}
=
\left(\ontop{2n}{n}\right)^2\;\frac{n^3(n\!-\!1)(7n^2-18n+13)}{60(2n-1)(2n-3)}\; .
\end{eqnarray*}},
\begin{eqnarray}
&&{\cal P}_2(n_1,n_2)=\frac{n_1n_2}{3}\; ,\nonumber\\
&&{\cal P}_4(n_1,n_2)=\frac{n_1 n_2\big({7 n_1n_2}
\!-\!({n_1\!+\!n_2})
\big)}{15}\; .
\label{eq:poly}
\end{eqnarray}
(The polynomials that enter in eq.~(\ref{eq:moments-0}) up to $2l\le 12$ are
listed at the end of the section \ref{sec:moments}.)

The rest of this paper is devoted to a proof of
eqs.~(\ref{eq:moments-0}) and (\ref{eq:poly}). Our approach is based
on the formulation of lattice scalar quantum electrodynamics in the
worldline formalism. When this formulation is used in two dimensions,
in the presence of a transverse magnetic field, the propagator of the
scalar particle can be viewed as a generating function for the
distribution of the areas of random loops on a square lattice. In the
section \ref{sec:QED}, we define more precisely the model and recall
this correspondence. We also explain how to perform an expansion in
powers of the interactions with the magnetic field. In the section
\ref{sec:moments}, we prove the main result, namely
eq.~(\ref{eq:moments-0}). Our proof leads to an algorithm to obtain
explicit expressions for the polynomial ${\cal P}_{2l}$, that we have
implemented in order to obtain the formulas (\ref{eq:poly}) and
several higher moments. A few intermediate calculations are explained
in the appendix \ref{app:combi}, and some basic links between our
approach and the Hofstadter-Harper Hamiltonian are outlined in the
appendix \ref{app:HHH}.

\section{Lattice scalar QED in two dimensions}
\label{sec:QED}
\subsection{Area distribution of random walks from lattice QED}
Consider the following operator, acting on complex valued functions on
${\mathbbm Z}^2$~:
\begin{eqnarray}
&&
\sum_{(k,l)\in{\mathbbm Z}^2}D_{ij,kl}\;f_{kl}\equiv 
\frac
{2f_{ij}-U_{1,ij}f_{i+1 j}-U_{1,i-1j}^*f_{i-1 j}}
{a_1^2}
\nonumber\\
&&\qquad\qquad\qquad
+\frac
{2f_{ij}-U_{2,ij}f_{i j+1}-U_{2,ij-1}^*f_{i j-1}}
{a_2^2}\; .
\label{eq:invprop-landau}
\end{eqnarray}
In Quantum Electrodynamics, this operator is the square of the
discrete 2-dimensional covariant derivative\footnote{When $U_{1,ij}=1$
  and $U_{2,ij}=1$, this operator is just the opposite of the discrete
  Laplacian.} in two dimensions, acting on a scalar field. Here, we
choose an anisotropic lattice, with lattice spacings $a_1$ and $a_2$
in the two directions. The coefficients $U_{1,ij}$ and $U_{2,ij}$ are
complex phases called ``link variables''. Their name comes from the
fact that they live on the edges of the lattice, while the function
$f_{ij}$ lives on the nodes of the lattice. The link variables have a
direction and an orientation: in the notation $U_{1,ij}$, the index
$1$ is the direction of the link, and the coordinates $ij$ denote the
starting point of the oriented link. Thus, $U_{1,ij}$ lives on the
edge that connects the points $(i,j)$ and $(i+1,j)$. The complex
conjugate of a link variable lives on the same link but has the
opposite orientation.

An interesting object is the inverse $G_{ij,kl}$ of this operator,
which in QED is nothing but the propagator of a particle in a given
external electromagnetic field. It is defined by
\begin{equation}
\sum_{(k,l)\in{\mathbbm Z}^2}D_{ij,kl}\;G_{kl;mn}=\delta_{im}\delta_{jn}\; .
\end{equation}
In the worldline formalism, this inverse can be written as a sum over
closed random walks (see ref.~\cite{EpelbGW2}). When evaluated with
identical endpoints (we take the point of coordinates $(0,0)$, but
this choice is irrelevant), its expression reads
\begin{equation}
G_{00,00}
=
\frac{{\bf a}^2}{4}
\sum_{n_1,n_2=0}^\infty
\left(\frac{h_1}{4}\right)^{2n_1}
\left(\frac{h_2}{4}\right)^{2n_2}
\sum_{\gamma\in{\bs\Gamma}_{n_1,n_2}}
\prod_{\ell\in\gamma}U_\ell
\; ,
\label{eq:G00}
\end{equation}
where $\prod_{\ell\in\gamma} U_\ell$ denotes the product of the link
variables encountered along the closed path $\gamma$.
We have also defined
\begin{equation}
\frac{2}{{\bf a}^2}\equiv\frac{1}{a_1^2}+\frac{1}{a_2^2}\quad,\qquad 
h_{1,2}\equiv \frac{{\bf a}^2}{a_{1,2}^2}\; .
\label{eq:rescaled}
\end{equation}

By using an anisotropic lattice, we can isolate closed random walks
with a given number of steps in each direction by selecting a given
term in the Taylor series of the left hand side in powers of $h_1$ and
$h_2$.  In other words, the order in $h_1$ and $h_2$ counts the
numbers of steps made by the random walk in the directions $\pm 1$ and
$\pm 2$. In order to obtain an explicit connection to the area
enclosed by the paths, we need to specialize the link variables in
such a way that they represent a magnetic field transverse to the
plane of the lattice. By Gauss's law, the product
$\prod_{\ell\in\gamma} U_\ell$ will then become the exponential of the
magnetic flux through the loop $\gamma$, and will give direct access
to the area. For a given transverse magnetic field $B$, there are in
fact infinitely many ways to represent it in terms of link variables,
due to the gauge invariance of electrodynamics\footnote{Different
  gauge choices correspond to different ways of writing the algebraic
  area inside a closed loop $\gamma$ as a contour integral. The Landau
  gauge used in this paper amounts to using the formula $\oint_\gamma
  x\,dy$.  Gauge transformations change the integrand into
  $x\,dy+d\theta(x,y)$, which leaves the area unchanged since the
  added term is a total derivative.}. The choice that we adopt in the
rest of this paper is Landau gauge~:
\begin{equation}
U_{1,ij}\equiv 1\quad,\qquad U_{2,ij}\equiv e^{iBa_1 a_2 i}\; .
\label{eq:gauge}
\end{equation}
It is easy to check that
\begin{equation}
\prod_{\ell\in\gamma}U_\ell = e^{iB a_1 a_2\,{\rm Area}\,(\gamma)}\; ,
\end{equation}
where ${\rm Area}\,(\gamma)$ is the algebraic area enclosed by the
path $\gamma$, measured as a number of elementary lattice
plaquettes. Note that this equation would be true with any other gauge
equivalent choice for the link variables\footnote{This is trivially
  seen by recalling that gauge transformations of the link variables
  read\begin{equation*} U_{1,ij}\to
    \Omega_{ij}U_{1,ij}\Omega^*_{i+1j}\quad,\qquad U_{2,ij}\to
    \Omega_{ij}U_{2,ij}\Omega^*_{ij+1}\; .\end{equation*}}, since the
left hand side is a gauge invariant quantity. In the rest of the
paper, we denote $\phi\equiv Ba_1a_2$ the magnetic flux through one
elementary plaquette of the lattice.

Knowing the left hand side of eq.~(\ref{eq:G00}) as a function of
$\phi,h_{1,2}$ would therefore give access to the full distribution of
areas, since we can write
\begin{equation}
\sum_{\gamma\in{\bs\Gamma}_{n_1,n_2}}
e^{i\phi\, {\rm Area}\,(\gamma)}
=
\sum_{l=0}^\infty \frac{(-1)^l}{(2l)!}\;\phi^{2l}
\sum_{\gamma\in{\bs\Gamma}_{n_1,n_2}}\left({\rm Area}\,(\gamma)\right)^{2l}\; .
\label{eq:moments}
\end{equation}
Only the even moments are non-zero, since one gets the opposite
algebraic area by reversing the path $\gamma$. Note that this
correspondence is equivalent to the well-known relationship between
the distribution of areas and the traces of the powers of the
Hofstadter-Harper Hamiltonian~\cite{BeguiVZ1,BelliCBC1,MashkO1}.

\subsection{Rules for the expansion in powers of the magnetic field}
\label{sec:rules}
In terms of the rescaled lattice spacings defined in
eq.~(\ref{eq:rescaled}), the operator $D$ reads
\begin{equation}
{\bf a}^2\,
\sum_{(k,l)\in{\mathbbm Z}^2}D_{ij,kl}\;f_{kl}
\equiv
4 f_{ij}
-h_1(f_{i+1 j}+f_{i-1 j})
-h_2(e^{i\phi i}f_{ij+1}+e^{-i\phi i}f_{ij-1})\; ,
\end{equation}
where we have now specialized to the gauge choice of
eq.~(\ref{eq:gauge}). This operator can be separated into a vacuum
part and a term due to the interactions with the magnetic field,
\begin{equation}
D=D^{(0)}-\Delta\; ,
\label{eq:D-delta}
\end{equation}
with
\begin{eqnarray}
{\bf a}^2\,\sum_{(k,l)\in{\mathbbm Z}^2}D^{(0)}_{ij,kl}\;f_{kl}
&\equiv&
4f_{ij}
-h_1(f_{i+1j}+f_{i-1j})
-h_2(f_{ij+1}+f_{ij-1})
\nonumber\\
{\bf a}^2\,\sum_{(k,l)\in{\mathbbm Z}^2}\Delta_{ij,kl}\;f_{kl}
&\equiv&
h_2((e^{i\phi i}-1)f_{ij+1}+(e^{-i\phi i}-1)f_{ij-1})\; .
\end{eqnarray}

Let us first consider the vacuum case, in order to establish the
notations. We consider an infinite lattice. In this case,
the propagator can be conveniently represented in Fourier space
by the following formula,
\begin{equation}
G^{(0)}_{ij,kl}
={\bf a}^2
\int_0^{2\pi}
\frac{dp_1 dp_2}{(2\pi)^2}\;
\frac{e^{i(p_1(i-k)+p_2(j-l))}}{4-2(h_1\,\cos(p_1)+h_2\,\cos(p_2))}\; .
\label{eq:prop-vac}
\end{equation}
By evaluating this expression with equal endpoints, $i=k,j=l$, and
expanding in powers of $h_{1,2}$, we can easily recover
eq.~(\ref{eq:G00}) for a vanishing flux $\phi=0$. For this, we only
need to recall the number of random loops that make $n_1$ hops in the
direction $+1$ and $n_2$ hops in the direction $+2$~:
\begin{equation}
\sum_{\gamma\in{\bs\Gamma}_{n_1,n_2}} 1
=
\frac{(2(n_1+n_2))!}{n_1!^2 n_2!^2}\; .
\end{equation}

In order to go beyond this trivial result, we must include the effect
of the interaction term $\Delta$.  The inverse of
eq.~(\ref{eq:D-delta}) can be written as a formal series
\begin{eqnarray}
G
=
G^{(0)}
+
\underbrace{G^{(0)}\Delta G^{(0)}}_{G^{(1)}}
+
\underbrace{G^{(0)}\Delta G^{(0)}\Delta G^{(0)}}_{G^{(2)}}
+\cdots=
\sum_{k=0}^\infty G^{(k)}\; .
\end{eqnarray}
Note that this is not exactly an expansion in powers of $\phi$, since
$\Delta$ is itself an infinite series in $\phi$. However, since
$\Delta$ starts at the order $\phi$, we need only to calculate the
terms up to the order $\Delta^{2l}$ if we are interested in the order
$\phi^{2l}$.

Note that, although the free propagator depends on a single momentum
$\p\equiv(p_1,p_2)$, the full propagator carries different momenta at
its two endpoints, because of the interaction with the magnetic field.
Thus, eq.~(\ref{eq:prop-vac}) is replaced by
\begin{equation}
G_{ij,kl}
=
\int_0^{2\pi}
\frac{dp_1 dp_2}{(2\pi)^2}\frac{dp_1' dp_2'}{(2\pi)^2}\;
e^{i(p_1i-p_1'k+p_2j-p_2'l)}\;\widetilde{G}(\p,\p')\; ,
\label{eq:prop}
\end{equation}
where $\widetilde{G}(\p,\p')$ is the full propagator in momentum
space. Here, we define the momenta in such a way that $\p$ {enters} at
one endpoint and $\p'$ {exits} at the other endpoint of the
propagator. The off-diagonal momentum components of the propagator are
inherited from those of the interaction term,
\begin{eqnarray}
{\bf a}^2\,\widetilde{\Delta}(\p,\p')
&=&
(2\pi)^2 h_2\Big[
e^{ip_2}\delta(p_1'-p_1-\phi)
+
e^{-ip_2}\delta(p_1'-p_1+\phi)
\nonumber\\
&&\qquad\qquad
-2\cos(p_2)\delta(p_1'-p_1)
\Big]\;\delta(p_2'-p_2)\; .
\label{eq:delta-F}
\end{eqnarray}
With the gauge choice that we have adopted, the interactions with the
magnetic field always conserve the component $p_2$ of the
momentum. Therefore, the full propagator $\widetilde{G}(\p,\p')$ is
itself proportional to $\delta(p_2'-p_2)$. In contrast, the magnetic
field can change the component $p_1$, but only in discrete increments
$\pm\phi$ (or 0 if we pick the third term in eq.~(\ref{eq:delta-F})).
For bookkeeping purposes, it is useful to write $\widetilde{\Delta}$
as a sum of three terms,
\begin{eqnarray}
&&\widetilde{\Delta}\equiv\sum_{\epsilon\in\{-1,0,+1\}}\widetilde{\Delta}_\epsilon
\nonumber\\
&&{\bf a}^2\,\widetilde{\Delta}_\epsilon \equiv (2\pi)^2\;h_2\;
\delta(p_1'-(p_1+\epsilon\phi))\;\delta(p_2'-p_2)\;V_\epsilon(p_2)
\nonumber\\
&&V_{\pm 1}(p_2)\equiv e^{\pm i p_2}\quad,\qquad V_0(p_2)\equiv -2 \cos(p_2)
\; .
\end{eqnarray}
The index $\epsilon$ denotes the increment (in units of $\phi$) of the
momentum $p_1$ caused by the interaction. Note that
\begin{equation}
V_{-1}+V_0+V_{+1}=0\; ,
\end{equation}
which implies that each insertion of the interaction term increases
the order in $\phi$ by at least one unit.

\subsection{Term of order $\Delta^k$}
We can organize
the expression of $\wt{G}^{(k)}$ as follows,
\begin{eqnarray}
\wt{G}^{(k)}
&=&
\sum_{(\epsilon_1,\cdots,\epsilon_k)\in\{-1,0,+1\}^k}
\wt{G}^{(k;\epsilon_1\cdots\epsilon_k)}\nonumber\\
\wt{G}^{(k;\epsilon_1\cdots\epsilon_k)}
&\equiv&
\wt{G}^{(0)}\widetilde{\Delta}_{\epsilon_1}\wt{G}^{(0)}\widetilde{\Delta}_{\epsilon_2}
\cdots\widetilde{\Delta}_{\epsilon_k}\wt{G}^{(0)}\; ,
\end{eqnarray}
where $\wt{G}^{(0)}$ is the Fourier transform of the free propagator,
that can be read off from eq.~(\ref{eq:prop-vac}). Tracking explicitly the
momentum flow, this can be written as
\begin{equation}
\wt{G}^{(k;\epsilon_1\cdots\epsilon_k)}
=
h_2^k\;\wt{G}^{(0)}(\p)\;V_{\epsilon_1}(p_2)\;\wt{G}^{(0)}(\p+\sigma_1{\bs \phi})\cdots V_{\epsilon_k}(p_2)\;\wt{G}^{(0)}(\p+\sigma_k{\bs \phi})\; ,
\end{equation}
where we denote
\begin{equation}
  {\bs\phi}\equiv(\phi,0)\quad,\qquad \sigma_i\equiv\sum_{j=1}^i\epsilon_j\qquad(\mbox{and\ }\sigma_0\equiv 0)\; .
  \label{eq:sigma}
\end{equation}
The contribution of this term to the left hand side of
eq.~(\ref{eq:G00}) reads
\begin{eqnarray}
&&
G^{(k;\epsilon_1\cdots\epsilon_k)}_{00,00}
=\frac{{\bf a}^2}{4}\int\frac{dp_1dp_2}{(2\pi)^2}\;
\frac{4\,h_2^k}{4-2(h_1\cos(p_1)+h_2\cos(p_2))}\nonumber\\
&&\!\times
\frac{V_{\epsilon_1}(p_2)}{4\!-\!2(h_1\cos(p_1\!+\!\sigma_1\phi)\!+\!h_2\cos(p_2))}
\cdots
\frac{V_{\epsilon_k}(p_2)}{4\!-\!2(h_1\cos(p_1\!+\!\sigma_k\phi)\!+\!h_2\cos(p_2))}
\; .
\nonumber\\
&&
\end{eqnarray}
In order to make the connection with eq.~(\ref{eq:G00}), let us expand
the integrand in powers of $h_{1,2}$~:
\begin{eqnarray}
&&\mbox{Integrand}=
{(-1)^{n_0}}
\sum_{\ontop{(a_0,\cdots,a_k,}{b_0,\cdots,b_k)\in{\mathbbm N}^{2k+2}}}
\frac{(a_0+b_0)!}{a_0!b_0!}\cdots\frac{(a_k+b_k)!}{a_k!b_k!}
\nonumber\\
&&\qquad\times\;
\left(\frac{h_1}{4}\right)^{a_0+\cdots+a_k}
\left(\frac{h_2}{4}\right)^{k+b_0+\cdots+b_k}
\!\!\big(2\cos(p_2)\big)^{b_0+\cdots+b_k+n_0}
\;e^{i(n_+-n_-)p_2}
\nonumber\\
&&\qquad\times\;
\big(2\cos(p_1)\big)^{a_0}
\big(2\cos(p_1+\sigma_1\phi)\big)^{a_1}
\cdots
\big(2\cos(p_1+\sigma_k\phi)\big)^{a_k}\; ,
\nonumber\\
&&
\label{eq:integrand}
\end{eqnarray}
where we denote
\begin{eqnarray}
n_{-,0,+}\equiv{\rm Card}\,\left\{1\le i\le k\big|\epsilon_i=-1,0,+1\right\}\; .
\label{eq:nnn}
\end{eqnarray}
(Note that $n_-+n_0+n_+=k$.) The integral over $p_2$ is of the form
\begin{equation}
\int_0^{2\pi}\frac{dp_2}{2\pi}\;
(2\cos p_2)^m\;e^{inp_2}
=\sum_{p+q=m}\frac{m!}{p!q!}\;\delta_{n,p-q}
=\frac{m!}{\big(\tfrac{m+n}{2}\big)!\big(\tfrac{m-n}{2}\big)!}\; .
\end{equation}
The final expression is valid only if $|n|\le m$ and $m,n$ have the
same parity, otherwise the integral is zero. Therefore, $m+n$ and
$m-n$ should both be even. Since $m=n_0+\sum b_i$ and $n=n_+-n_-$,
this implies that $k+\sum b_i$ should be even. This was of course
expected for a closed random walk, since $k+\sum b_i$ is the number of
steps in the directions $\pm 2$ (see the order in $h_2$ in
eq.~(\ref{eq:integrand})). For the integral over $p_1$, we need
\begin{eqnarray}
&&\int_0^{2\pi}\frac{dp_1}{2\pi}\;
\big(2\cos(p_1+\sigma_0\phi)\big)^{a_0}\cdots
\big(2\cos(p_1+\sigma_k\phi)\big)^{a_k}
=
\sum_{\alpha_0+\beta_0=a_0}\cdots
\nonumber\\
&&\cdots\!\!\!\!\sum_{\alpha_k+\beta_k=a_k}
\!\!
\frac{a_0!}{\alpha_0!\beta_0!}\cdots\frac{a_k!}{\alpha_k!\beta_k!}\,
\delta_{\alpha_0+\cdots+\alpha_k,\beta_0+\cdots+\beta_k}\,
e^{i\phi\big(\sigma_0(\alpha_0-\beta_0)+\cdots+\sigma_k(\alpha_k-\beta_k)\big)}
\nonumber\\
&&
\end{eqnarray}
Wrapping everything together, we obtain
\begin{eqnarray}
&&
G^{(k;\epsilon_1\cdots\epsilon_k)}_{00,00}
=\frac{{\bf a}^2}{4}{(-1)^{n_0}}
\sum_{n_1,n_2}
\sum_{\ontop{\alpha_0+\cdots+\alpha_k=n_1}{\ontop{\beta_0+\cdots+\beta_k=n_1}{k+b_0+\cdots+b_k=2n_2}}}
\!\!\!\!\!\!\!\!\!\!
\frac{(\alpha_0+\beta_0+b_0)!}{\alpha_0!\beta_0!b_0!}\cdots
\frac{(\alpha_k+\beta_k+b_k)!}{\alpha_k!\beta_k!b_k!}
\nonumber\\
&&\quad\times\;
\left(\frac{h_1}{4}\right)^{2n_1}\!\!
\left(\frac{h_2}{4}\right)^{2n_2}\!\!\!
\frac{(2n_2-n_+-n_-)!}{\big(n_2-n_+\big)!\big(n_2-n_-\big)!}\;
e^{i\phi\big(\sigma_0(\alpha_0-\beta_0)+\cdots+\sigma_k(\alpha_k-\beta_k)\big)}\, ,
\nonumber\\
&&
\label{eq:pert-1}
\end{eqnarray}
where we have introduced
\begin{equation}
2n_1\equiv \sum_{i=0}^ka_i\quad,\qquad
2n_2\equiv k+\sum_{i=0}^kb_i\; .
\end{equation}
The sum over the $b_i$'s can be performed easily by using the
following identity (see the appendix \ref{app:comb1})
\begin{equation}
\sum_{b_0+\cdots+b_k=B}\frac{(a_0+b_0)!}{b_0!}\cdots\frac{(a_k+b_k)!}{b_k!}
=
a_0!\cdots a_k!\;\frac{(A+B+k)!}{B!(A+k)!}\; ,
\label{eq:comb1}
\end{equation}
where $A\equiv a_0+\cdots+a_k$. Therefore, eq.~(\ref{eq:pert-1}) can
be rewritten as
\begin{equation}
G^{(k;\epsilon_1\cdots\epsilon_k)}_{00,00}
=
\frac{{\bf a}^2}{4}
\sum_{n_1,n_2}
\left(\!\frac{h_1}{4}\!\right)^{2n_1}\!
\left(\!\frac{h_2}{4}\!\right)^{2n_2}
\frac{(2(n_1\!+\!n_2))!}{n_1!^2n_2!^2}\;F_{k;\epsilon_1\cdots\epsilon_k}(n_1,n_2;\phi)
\; ,
\end{equation}
with
\begin{eqnarray}
&&F_{k;\epsilon_1\cdots\epsilon_k}(n_1,n_2;\phi)\equiv{(-1)^{n_0}}\;
\frac{n_1!^2}{(2n_1\!+\!k)!}\frac{n_2!^2}{(2n_2\!-\!k)!}\;
\frac{(2n_2\!-\!n_+\!-\!n_-)!}{\big(n_2\!-\!n_+\big)!\big(n_2\!-\!n_-\big)!}
\nonumber\\
&&\times
\!\sum_{\ontop{\alpha_0+\cdots+\alpha_k=n_1}{\beta_0+\cdots+\beta_k=n_1}}
\!\!\!\!\!
\frac{(\alpha_0\!+\!\beta_0)!}{\alpha_0!\beta_0!}\cdots
\frac{(\alpha_k\!+\!\beta_k)!}{\alpha_k!\beta_k!}\,
e^{i\phi\big(\sigma_0(\alpha_0-\beta_0)+\cdots+\sigma_k(\alpha_k-\beta_k)\big)}\, .
\nonumber\\
&&
\label{eq:pert-2}
\end{eqnarray}
In this formula, the sum over the $\alpha_i$'s and $\beta_i$'s, as
well as the combinatorial factors under this sum, are symmetric under
the exchange $\alpha_i\leftrightarrow\beta_i$, while the argument of
the exponential is antisymmetric. Therefore,
$F_{k;\epsilon_1\cdots\epsilon_k}(n_1,n_2;\phi)$ is even in $\phi$.

\section{Moments of the distribution of areas}
\label{sec:moments}
\subsection{Terms of order $\phi^0$}
As a check, one can first evaluate eq.~(\ref{eq:pert-2}) for
$\phi=0$. For this, we need the following sum (see the appendix \ref{app:comb2})
\begin{equation}
\sum_{\ontop{\alpha_0+\cdots+\alpha_k=n_1}{\beta_0+\cdots+\beta_k=n_1}}
\!\!\!\!\!
\frac{(\alpha_0\!+\!\beta_0)!}{\alpha_0!\beta_0!}\cdots
\frac{(\alpha_k\!+\!\beta_k)!}{\alpha_k!\beta_k!}
=\frac{(2n_1+k)!}{k!n_1!^2}\; ,
\label{eq:comb2}
\end{equation}
which leads to 
\begin{equation}
F_{k;\epsilon_1\cdots\epsilon_k}(n_1,n_2;0)
=
\frac{(-1)^{n_0}}{k!}
\frac{n_2!^2}{(2n_2\!-\!k)!}\;
\frac{(2n_2\!-\!n_+\!-\!n_-)!}{\big(n_2\!-\!n_+\big)!\big(n_2\!-\!n_-\big)!}\; .
\end{equation}
When summing over all the possible assignments of $\epsilon_i$ in
$\{-1,0,+1\}$, we obtain
\begin{equation}
\frac{n_2!^2}{k!(2n_2-k)!}
\sum_{\ontop{(\epsilon_1,\cdots,\epsilon_k)}{\in\{-1,0,+1\}^k}}
(-1)^{n_0}
\frac{(2n_2\!-\!n_+\!-\!n_-)!}
{\big(n_2\!-\!n_+\big)!\big(n_2\!-\!n_-\big)!}=0{\rm \ \ if\ }k\ge 1\; ,
\label{eq:comb3}
\end{equation}
(see the appendix \ref{app:comb3}) that vanishes as expected since the
interaction terms do not contribute to the $\phi$-independent part of
the result.

\subsection{Terms of order $\phi^2$}
In order to obtain the terms that are quadratic in the magnetic flux,
we need to calculate the sum
\begin{eqnarray}
&&
\sum_{\ontop{\alpha_0+\cdots+\alpha_k=n_1}{\beta_0+\cdots+\beta_k=n_1}}
\!\!\!\!\!
\frac{(\alpha_0\!+\!\beta_0)!}{\alpha_0!\beta_0!}\cdots
\frac{(\alpha_k\!+\!\beta_k)!}{\alpha_k!\beta_k!}
\;
\big[\sigma_1(\alpha_1-\beta_1)+\cdots+\sigma_k(\alpha_k-\beta_k)\big]^2
=
\nonumber\\
&&=\!\!\!
\sum_{\ontop{\alpha_0+\cdots+\alpha_k=n_1}{\beta_0+\cdots+\beta_k=n_1}}
\!\!\!\!\!
\frac{(\alpha_0\!+\!\beta_0)!}{\alpha_0!\beta_0!}\cdots
\frac{(\alpha_k\!+\!\beta_k)!}{\alpha_k!\beta_k!}
\;
\Big[\sum_{i=1}^k\sigma_i^2(\alpha_i-\beta_i)^2
\nonumber\\
&&\qquad\qquad\qquad\qquad\qquad\qquad\qquad\qquad
+\smash{
\sum_{i\not=j}\sigma_i\sigma_j(\alpha_i-\beta_i)(\alpha_j-\beta_j)\Big]
}
\; .\nonumber\\
&&
\end{eqnarray}
The terms that appear in the right hand side involve sums of one of
the following types (see the appendices \ref{app:comb4} and \ref{app:comb5}),
\begin{equation}
\sum_{\ontop{\alpha_0+\cdots+\alpha_k=n_1}{\beta_0+\cdots+\beta_k=n_1}}
\!\!\!\!\!
\frac{(\alpha_0\!+\!\beta_0)!}{\alpha_0!\beta_0!}\cdots
\frac{(\alpha_k\!+\!\beta_k)!}{\alpha_k!\beta_k!}
\;
(\alpha_i-\beta_i)^2
=
\frac{2kn_1}{(k+2)!}\frac{(2n_1+k)!}{n_1!^2}\; ,
\label{eq:comb4}
\end{equation}
\begin{equation}
\sum_{\ontop{\alpha_0+\cdots+\alpha_k=n_1}{\beta_0+\cdots+\beta_k=n_1}}
\!\!\!\!\!
\frac{(\alpha_0\!+\!\beta_0)!}{\alpha_0!\beta_0!}\cdots
\frac{(\alpha_k\!+\!\beta_k)!}{\alpha_k!\beta_k!}
\;
(\alpha_i-\beta_i)(\alpha_j-\beta_j)
=
-\frac{2n_1}{(k+2)!}\frac{(2n_1+k)!}{n_1!^2}\; ,
\label{eq:comb5}
\end{equation}
($i\not=j$ in the second equation) and we
obtain\begin{eqnarray} &&
  \sum_{\ontop{\alpha_0+\cdots+\alpha_k=n_1}{\beta_0+\cdots+\beta_k=n_1}}
  \!\!\!\!\!  \frac{(\alpha_0\!+\!\beta_0)!}{\alpha_0!\beta_0!}\cdots
  \frac{(\alpha_k\!+\!\beta_k)!}{\alpha_k!\beta_k!}  \;
  \big[\sigma_1(\alpha_1-\beta_1)+\cdots+\sigma_k(\alpha_k-\beta_k)\big]^2
  =
  \nonumber\\
  &&\qquad\qquad=
  \frac{(2n_1+k)!}{n_1!^2}\frac{2n_1}{(k+2)!}\;
\Big[
k\sum_{i=1}^k\sigma_i^2
-\sum_{i\not=j}\sigma_i\sigma_j
\Big]\; .
\end{eqnarray}
Since the expansion in $\Delta^k$ is in fact an expansion in powers of
$e^{\pm i\phi}-1$, it is a priori obvious that the terms of order
$\phi^2$ can only appear
in the orders $k=1$ and $k=2$. This corresponds to a fairly small
number of terms ($3$ for $k=1$ and $3^2$ for $k=2$) that can be summed
by hand to obtain
\begin{eqnarray}
F_{k=0}(n_1,n_2;\phi)&=&1
\nonumber\\
\sum_{\epsilon_1\in\{-1,0,+1\}}\!\!\!\!\!F_{k=1;\epsilon_1}(n_1,n_2;\phi)&=&
-\frac{\phi^2}{2!}
\;
\frac{4n_1n_2}{3!}+{\cal O}(\phi^4)
\nonumber\\
\sum_{(\epsilon_1,\epsilon_2)\in\{-1,0,+1\}^2}\!\!\!\!\!F_{k=2;\epsilon_1\epsilon_2}(n_1,n_2;\phi)&=&
+\frac{\phi^2}{2!}
\;
\frac{8n_1n_2}{4!}+{\cal O}(\phi^4)\; ,
\end{eqnarray}
so that the total up to the order $\phi^2$ reads
\begin{equation}
\sum_{k\le2}\;\sum_{(\epsilon_1,\cdots,\epsilon_k)\in\{-1,0,+1\}^k}\!\!\!\!\!F_{k;\epsilon_1\cdots\epsilon_k}(n_1,n_2;\phi)
=
1-\frac{\phi^2}{2!}\;\frac{n_1n_2}{3}+{\cal O}(\phi^4)\; .
\end{equation}
From this formula, we can read out the second moment of the
distribution of areas,
\begin{equation}
\sum_{\gamma\in{\bs\Gamma}_{n_1,n_2}}
\!\!\!\!\!\!
\left({\rm Area}\,(\gamma)\right)^{2}
=\frac{(2(n_1\!+\!n_2))!}{n_1!^2n_2!^2}\;\frac{n_1n_2}{3}\; ,
\end{equation}
which ends the proof of the first of eqs.~(\ref{eq:poly}).

\subsection{Moment of order ${2l}$}
\subsubsection{General structure}
Let us now consider the general case of the terms of arbitrary (but
even) order $\phi^{2l}$. Now, the sum we must calculate is
\begin{eqnarray}
&&
\sum_{\ontop{\alpha_0+\cdots+\alpha_k=n_1}{\beta_0+\cdots+\beta_k=n_1}}
\!\!\!\!\!
\frac{(\alpha_0\!+\!\beta_0)!}{\alpha_0!\beta_0!}\cdots
\frac{(\alpha_k\!+\!\beta_k)!}{\alpha_k!\beta_k!}
\;
\big[\sigma_1(\alpha_1-\beta_1)+\cdots+\sigma_k(\alpha_k-\beta_k)\big]^{2l}
=
\nonumber\\
&&\!\!\!\!\!\!\!\!\!\!=\!\!\!\!\!
\sum_{l_1+\cdots+l_k=2l}\frac{(2l)!}{l_1!\cdots l_k!}\;
\sigma_1^{l_1}\cdots\sigma_k^{l_k}\!\!
\underbrace{\sum_{\ontop{\alpha_0+\cdots+\alpha_k=n_1}{\beta_0+\cdots+\beta_k=n_1}}
\!\!\!\!\!
\frac{(\alpha_0\!+\!\beta_0)!}{\alpha_0!\beta_0!}\cdots
\frac{(\alpha_k\!+\!\beta_k)!}{\alpha_k!\beta_k!}
\prod_{i=1}^k(\alpha_i\!-\!\beta_i)^{l_i}}_{{\cal C}_{l_1\cdots l_k}(n_1)}
.
\nonumber\\
&&
\label{eq:tmp4}
\end{eqnarray}
In order to evaluate the sum underlined in the last line, we proceed
as in the appendix \ref{app:comb4}. First, we define a generating
function
\begin{equation}
H_{l_1\cdots l_k}(x,y)\equiv\sum_{n_1,n_1'\ge 0}x^{n_1}y^{n_1'}\!\!\!
\sum_{\ontop{\alpha_0+\cdots+\alpha_k=n_1}{\beta_0+\cdots+\beta_k=n_1'}}
\!\!\!\!\!
\frac{(\alpha_0\!+\!\beta_0)!}{\alpha_0!\beta_0!}\cdots
\frac{(\alpha_k\!+\!\beta_k)!}{\alpha_k!\beta_k!}\; \prod_{i=1}^k(\alpha_i-\beta_i)^{l_i}\; .
\end{equation}
The desired quantity is the coefficient of $(xy)^{n_1}$ in the
Taylor expansion of this function. Following the appendices
\ref{app:comb4} and \ref{app:comb5}, this function is also equal to
\begin{equation}
H_{l_1\cdots l_k}(x,y)
=\frac{1}{1-x-y}\prod_{i=1}^{k}
\left[(x\partial_x-y\partial_y)^{l_i}\frac{1}{1-x-y}\right]\; .
\end{equation}
After evaluating the derivatives, the generating function takes the
form
\begin{equation}
H_{l_1\cdots l_k}(x,y)=\frac{A_{l_1}(x,y)A_{l_2}(x,y)\cdots A_{l_k}(x,y)}{(1-x-y)^{1+k+2l}}\; ,
\label{eq:gen-func}
\end{equation}
where the $A_n(x,y)$ are polynomials defined iteratively by
\begin{eqnarray}
A_0(x,y)&=&1\nonumber\\
A_{n+1}(x,y)&=&
\Big[(n+1)(x\!-\!y)+(1\!-\!x\!-\!y)(x\partial_x\!-\!y\partial_y)\Big]\,A_n(x,y)\; .
\end{eqnarray}
The degree of $A_n$ is thus equal to $n$, and the product in the
numerator of eq.~(\ref{eq:gen-func}) can be expanded as
\begin{equation}
\prod_{i=1}^k A_{l_i}(x,y)=\sum_{m+n\le 2l} A_{mn}\big(\{l_i\}\big)\;x^m y^n\; .
\end{equation}
Each term in this sum gives the following contribution to the
coefficient of $(xy)^{n_1}$ in the Taylor expansion of the generating
function~:
\begin{equation}
\frac{x^m y^n}{(1-x-y)^{1+k+2l}}=\cdots
+
\frac{(2n_1+k+2l-m-n))!}{(k+2l)!\,(n_1-m)!(n_1-n)!}\;(xy)^{n_1}+\cdots\; .
\end{equation}
Let us combine ${\cal C}_{l_1\cdots l_k}(n_1)$ with the other factors
that depend on $n_1$ in eq.~(\ref{eq:pert-2}), at the exception of the
factor $(2(n_1+n_2))!/(n_1!^2n_2!^2)$,
\begin{eqnarray}
  &&P_{l_1\cdots l_k}^{(k)}(n_1)
\equiv
\frac{n_1!^2}{(2n_1\!+\!k)!}\,{\cal C}_{l_1\cdots l_k}(n_1)
\nonumber\\
&&\quad\;
=
\frac{1}{(k\!+\!2l)!}\!\!\sum_{m+n\le 2l}\!\!A_{mn}\big(\{l_i\}\big)
\underbrace{\frac{n_1!}{(n_1\!-\!m)!}}_{{\rm deg.\ }m}
\underbrace{\frac{n_1!}{(n_1\!-\!n)!}}_{{\rm deg.\ }n}
\underbrace{\frac{(2n_1\!+\!k\!+\!2l\!-\!m\!-\!n))!}{(2n_1\!+\!k)!}}_{{\rm deg.\ }2l-m-n}\,.
\nonumber\\
&&
\label{eq:pol1}
\end{eqnarray}
The three underlined factors are polynomials in $n_1$, of respective
degrees $m$, $n$ and $2l-m-n$, and therefore $P_{l_1\cdots
  l_k}^{(k)}(n_1)$ is also a polynomial in $n_1$ whose degree is
bounded by $2l$ (at this level of the discussion, it is not possible
to see the cancellations that may reduce the final degree when we sum
on $m,n$).

Let us now focus on the dependence on $n_2$, and combine all the
$n_2$-dependent factors into the following quantity, except the factor
$(2(n_1+n_2))!/(n_1!^2n_2!^2)$,
\begin{equation}
  Q_{l_1\cdots l_k}^{(k)}(n_2)\equiv
  \sum_{\ontop{(\epsilon_1,\cdots,\epsilon_k)}
{\in\{-1,0,+1\}^k}}
  (-1)^{n_0}\Big[\prod_{i=1}^k
  \sigma_i^{l_i}\Big]
  \underbrace{\frac{n_2!}{(n_2\!-\!n_+)!}}_{{\rm deg.\ }n_+}
  \underbrace{\frac{n_2!}{(n_2\!-\!n_-)!}}_{{\rm deg.\ }n_-}
  \underbrace{\frac{(2n_2\!-k\!+\!n_0)!}{(2n_2\!-\!k)!}}_{{\rm deg.\ }n_0}
\;,
\label{eq:pol2}
\end{equation}
which is a sum of products of three polynomials in $n_2$, of
respective degrees $n_+$, $n_-$ and $n_0$ (the total degree is thus
$k$). Therefore, $Q_{l_1\cdots l_k}^{(k)}(n_2)$ is a polynomial in
$n_2$ whose degree is bounded by $k$ (again, cancellations that may
decrease the degree when summing over the $\epsilon_i$'s are beyond
the reach of this argument -- see the subsection \ref{sec:deg-Qn2}).

In order to obtain the moment $2l$ of the distribution of areas, we
just need to sum\footnote{Note that
  $\sigma_1^{l_1}\cdots\sigma_k^{l_k}$ is a homogeneous polynomial of
  degree $2l$ in the $\epsilon_i$'s. If $k>2l$, some of the
  $\epsilon_i$'s must be absent in each monomial of this
  polynomial. By summing over the three values $-1,0,+1$ of any of the
  missing $\epsilon_i$'s, one gets zero (see the subsection
  \ref{sec:deg-Qn2}).} on $1\le k\le 2l$ and for each $k$ on the
partitions $l_1+\cdots+l_k=2l$. This leads to the following expression
for the $2l$-th moment~:
\begin{equation}
\sum_{\gamma\in{\bs\Gamma}_{n_1,n_2}}
\!\!\!\!\!\!
\left({\rm Area}\,(\gamma)\right)^{2l}
=\frac{(2(n_1\!+\!n_2))!}{n_1!^2n_2!^2}\;{\cal P}_{2l}(n_1,n_2)\; ,
\end{equation}
where ${\cal P}_{2l}$ is a polynomial in $n_1,n_2$ defined as~:
\begin{equation}
{\cal P}_{2l}(n_1,n_2)\equiv\sum_{\ontop{1\le k \le 2l}{l_1+\cdots+l_k=2l}}
\!\!\!\!\frac{(2l)!}{l_1!\cdots l_k!}\;
P_{l_1\cdots l_k}^{(k)}(n_1)\;Q_{l_1\cdots l_k}^{(k)}(n_2)\; .
\label{eq:pol3}
\end{equation}
The above counting only provides an upper bound $4l$ for the total
degree of this polynomial. However, the asymptotic behavior of these
moments for large random loops is known, since their area grows like
their perimeter: this implies that the degree should in fact be
exactly $2l$. In addition, this polynomial should obviously be
symmetric in $n_{1,2}$. This last property is non trivial in our
approach, since the choice of the gauge in which we have represented
the magnetic field breaks the manifest symmetry between the two
directions of space.

\subsubsection{Degree of $Q_{l_1\cdots l_k}^{(k)}(n_2)$}
\label{sec:deg-Qn2}
Starting from the expression (\ref{eq:pol2}) and expanding the factors
$\sigma_i^{l_i}$, we see that $Q_{l_1\cdots l_k}^{(k)}(n_2)$ is a
linear combination of the following polynomials~:
\begin{equation}
  S_{\lambda_1\cdots \lambda_k}^{(k)}(n_2)\equiv
  \sum_{\ontop{(\epsilon_1,\cdots,\epsilon_k)}
{\in\{-1,0,+1\}^k}}
  \!\!(-1)^{n_0}\Big[\prod_{i=1}^k
  \epsilon_i^{\lambda_i}\Big]
  {\frac{n_2!}{(n_2\!-\!n_+)!}}
  {\frac{n_2!}{(n_2\!-\!n_-)!}}
  {\frac{(2n_2\!-k\!+\!n_0)!}{(2n_2\!-\!k)!}}
\;,
\label{eq:pol2-1}
\end{equation}
where the $\lambda_i$ are integers $\ge 0$, such that
\begin{equation}
\sum_{i=1}^k\lambda_i=\sum_{i=1}^k l_i=2l\; .
\end{equation}
(This follows from the fact that the right hand side of
eq.~(\ref{eq:pol2}) is homogeneous in the $\epsilon_i$'s.)

Firstly, one can prove that $S_{\lambda_1\cdots
  \lambda_k}^{(k)}(n_2)=0$ if any of the $\lambda_i$'s is zero. Let us
assume for instance that $\lambda_k=0$, and introduce the following
notations~:
\begin{eqnarray}
\wt{n}_{-,0,+}\equiv{\rm Card}\,\left\{1\le i\le k-1\big|\epsilon_i=-1,0,+1\right\}\; ,
\end{eqnarray}
so that
\begin{eqnarray}
&&n_-=\wt{n}_-+1\;,\quad
n_{0,+}=\wt{n}_{0,+}\qquad\mbox{if\ }\epsilon_k=-1\nonumber\\
&&n_0=\wt{n}_0+1\;,\quad
n_{-,+}=\wt{n}_{-,+}\qquad\mbox{if\ }\epsilon_k=0\nonumber\\
&&n_+=\wt{n}_++1\;,\quad
n_{-,0}=\wt{n}_{-,0}\qquad\mbox{if\ }\epsilon_k=+1\; .
\end{eqnarray}
Then, the sum over the three values of $\epsilon_k$ contains a factor
\begin{eqnarray}
&&
\sum_{\epsilon_k\in\{-1,0,+1\}}
(-1)^{n_0}
\frac{(2n_2\!-\!n_+\!-\!n_-)!}
{\big(n_2\!-\!n_+\big)!\big(n_2\!-\!n_-\big)!}=\nonumber\\
&&
=(-1)^{\wt{n}_0}\frac{(2n_2-k+\wt{n}_0)!}{(n_2-\wt{n}_+)!(n_2-\wt{n}_-)!}
\Big\{
(n_2-\wt{n}_-)-(2n_2-\wt{n}_+-\wt{n}_-)\nonumber\\
&&\qquad\qquad\qquad\qquad\qquad\qquad\qquad\qquad\smash{+(n_2-\wt{n}_+)
\Big\}}=0\; .
\end{eqnarray}
(Notice that $\wt{n}_-+\wt{n}_0+\wt{n}_+=k-1$.)

From now on, we need only to consider the situation where all the
$\lambda_i$'s are strictly positive. First, note that since the
$\lambda_i$'s are non-zero, we can exclude $\epsilon_i=0$ from the sum
in eq.~(\ref{eq:pol2-1}). Let us assume that $k_{\rm e}$ of the
$\lambda_i$'s are even ($\ge 2$) and that $k_{\rm o}$ of them are odd
($k_{\rm e}+k_{\rm o}=k$). Without loss of generality, we can assume
that $\lambda_1,\cdots,\lambda_{k_{\rm o}}$ are odd and
$\lambda_{k_{\rm o}+1},\cdots,\lambda_{k}$ are even.  If $\lambda_k$
is even, it is easy to check that
\begin{equation}
 S_{\lambda_1\cdots \lambda_k}^{(k)}(n_2)
=
(2n_2-k+1)\;S_{\lambda_1\cdots \lambda_{k-1}}^{(k-1)}(n_2)
\end{equation}
by performing explicitly the sum over $\epsilon_k=\pm1$. By iterating
this formula, we can eliminate all the even $\lambda_i$'s to obtain
\begin{equation}
S_{\lambda_1\cdots \lambda_k}^{(k)}(n_2)
=
\frac{(2n_2-k+k_{\rm e})!}{(2n_2-k)!}\;
S_{\lambda_1\cdots \lambda_{k_{\rm o}}}^{(k_{\rm o})}(n_2)\; .
\label{eq:pol2-3}
\end{equation}

We need now to examine eq.~(\ref{eq:pol2-1}) in the case where all the
$\lambda_i$'s are odd, in order to evaluate the last factor in the
right hand side of the previous equation. When $\lambda$ is odd and
$\epsilon=\pm 1$, we have $\epsilon^\lambda=\epsilon$. Therefore,
\begin{equation}
\epsilon_1^{\lambda_1}\cdots \epsilon_{k_{\rm o}}^{\lambda_{\rm o}}
=\epsilon_1\cdots \epsilon_{k_{\rm o}}
=(-1)^{n_{{\rm o}-}}\; ,
\end{equation}
where we denote
\begin{equation}
n_{{\rm o}\pm}\equiv{\rm Card}\big\{1\le i\le k_{\rm o}\big|\epsilon_i=\pm 1\big\}\; .
\end{equation}
Thus, we have
\begin{equation}
  S_{\lambda_1\cdots \lambda_{k_{\rm o}}}^{(k_{\rm o})}(n_2)
=
  \!\!
  \sum_{(\epsilon_1,\cdots,\epsilon_{k_{\rm o}})\in\{-1,+1\}^{k_{\rm o}}}\!\!\!\!
  (-1)^{n_{{\rm o}-}}
  {\frac{n_2!}{(n_2\!-\!n_{{\rm o}+})!}}
  {\frac{n_2!}{(n_2\!-\!n_{{\rm o}-})!}}
\;.
\label{eq:pol2-odd}
\end{equation}
Since the right hand side depends on the $\epsilon_i$'s only via
$n_{{\rm o}\pm}$, we can rewrite this as
\begin{eqnarray}
  S_{\lambda_1\cdots \lambda_{k_{\rm o}}}^{(k_{\rm o})}(n_2)
&=&
  \sum_{n_{{\rm o}+}+n_{{\rm o}-}=k_{\rm o}}
  \frac{k_{\rm o}!}{n_{{\rm o}+}!n_{{\rm o}-}!}
  (-1)^{n_{{\rm o}-}}
  {\frac{n_2!}{(n_2\!-\!n_{{\rm o}+})!}}
  {\frac{n_2!}{(n_2\!-\!n_{{\rm o}-})!}}
\nonumber\\
&=&
\left\{
\begin{aligned}
&0&&\mbox{if } k_{\rm o}\mbox{ is odd}\\
&(-1)^p\frac{(2p)!}{p!}\frac{n_2!}{(n_2-p)!}&&\mbox{if } k_{\rm o}=2p\mbox{ is even}\\
\end{aligned}
\right.
\;.
\label{eq:pol2-odd-1}
\end{eqnarray}
Note that since $\sum_{i=1}^k\lambda_i=2l$ is even, the number $k_{\rm
  o}$ of odd $\lambda_i$'s is also even, and we are in the second
case.  By counting the powers of $n_2$ in eqs.~(\ref{eq:pol2-3}) and
(\ref{eq:pol2-odd-1}), we see that
\begin{equation}
{\rm deg}\,\left(S_{\lambda_1\cdots \lambda_k}^{(k)}\right)
=
k_{\rm e}+\frac{k_{\rm o}}{2}\; .
\end{equation}
Given our assumption that none of the $\lambda_i$'s is zero, this
leads to\footnote{If $k\le l$, the largest degree is realized when
  $\lambda_1,\cdots,\lambda_k$ are all even, so that $k_{\rm o}=0,
  k_{\rm e}=k$ and the degree is at most $k$ in this case.  Therefore,
  the degree of $S_{\lambda_1\cdots \lambda_k}^{(k)}$ is in fact
  bounded by ${\rm Min}\,(k,l)$.}
\begin{equation}
{\rm deg}\,\left(S_{\lambda_1\cdots \lambda_k}^{(k)}\right)
=\frac{k_{\rm o}}{2}+k_{\rm e}\le\frac{1}{2}\sum_{i=1}^k\lambda_i=l\; .
\end{equation}
Since the polynomial $Q_{l_1\cdots l_k}^{(k)}(n_2)$ is a linear
combination of the $S_{\lambda_1\cdots \lambda_k}^{(k)}(n_2)$, its
degree is at most $l$.  By symmetry, the maximum degree in the
variable $n_1$ in the moment of order $2l$ is also $l$.

\subsubsection{Explicit results for $2< 2l\le 12$}
The proof of the general structure of the moment of order $2l$ also
provides a straightforward algorithm to calculate explicitly the
polynomial ${\cal P}_{2l}(n_1,n_2)$. Let us mention an important
improvement over eq.~(\ref{eq:pol3}), based on the fact that the
polynomial $P_{l_1\cdots l_k}^{(k)}(n_1)$ is invariant under the
permutations of $\{l_1,\cdots,l_k\}$. One can therefore arrange the
partitions $\{l_1,\cdots,l_k\}$ into classes whose elements are
identical up to a permutation, and compute $P_{l_1\cdots
  l_k}^{(k)}(n_1)$ only once for each class. For each class, the sum
over the members of the class can be introduced inside the calculation
of $Q_{l_1\cdots l_k}^{(k)}(n_2)$, where it becomes the innermost
sum. For small $l$ up to $2l=12$, it leads to the following
expressions (we have arranged them in terms of the elementary
symmetric polynomials)~:
\begin{eqnarray}
&&{\cal P}_4(n_1,n_2)=\frac{n_1 n_2\big({7 n_1n_2}
\!-\!({n_1\!+\!n_2})
\big)}{15}\;,
\nonumber\\
&&{\cal P}_6(n_1,n_2)=\frac{n_1 n_2}{21}\big(31 (n_1n_2)^2
\!-\!15(n_1n_2)(n_1\!+\!n_2)
\nonumber\\
&&\qquad\qquad\qquad
+2(n_1\!+\!n_2)^2
\!-\!({n_1\!+\!n_2})
\big)\;,
\nonumber\\
&&{\cal P}_8(n_1,n_2)=\frac{n_1 n_2}{15}
\big(
127(n_1n_2)^3
\!-\!134(n_1n_2)^2(n_1\!+\!n_2)
+53(n_1n_2)(n_1\!+\!n_2)^2
\nonumber\\
&&\qquad\qquad\qquad
\!-\!6(n_1\!+\!n_2)^3
\!-\!22(n_1n_2)(n_1\!+\!n_2)
+8(n_1\!+\!n_2)^2
\!-\!3(n_1\!+\!n_2)
\big)\;,\nonumber\\
&&{\cal P}_{10}(n_1,n_2)=\frac{n_1 n_2}{33}
\big(
2555(n_1n_2)^4
\!-\!4778(n_1n_2)^3(n_1\!+\!n_2)
\nonumber\\
&&\qquad\qquad\qquad
+3745(n_1n_2)^2(n_1\!+\!n_2)^2
\!-\!1282(n_1n_2)(n_1\!+\!n_2)^3
\nonumber\\
&&\qquad\qquad\qquad
+120(n_1\!+\!n_2)^4
\!-\!1444(n_1n_2)(n_1\!+\!n_2)
+1438(n_1n_2)(n_1\!+\!n_2)^2
\nonumber\\
&&\qquad\qquad\qquad
\!-\!300(n_1\!+\!n_2)^3
\!-\!503(n_1n_2)(n_1\!+\!n_2)
+270(n_1\!+\!n_2)^2
\nonumber\\
&&\qquad\qquad\qquad
\!-\!85(n_1\!+\!n_2)
\big)\; ,
\label{eq:poly-1}
\end{eqnarray}
and
\begin{eqnarray}
&&{\cal P}_{12}(n_1,n_2)=
\frac{1414477(n_1n_2)^6}{1365}
-\frac{197569(n_1n_2)^5(n_1\!+\!n_2)}{65}
\nonumber\\
&&\qquad\qquad\qquad
+\frac{5381569(n_1n_2)^4(n_1\!+\!n_2)^2}{1365}
-\frac{2015366(n_1n_2)^4(n_1\!+\!n_2)}{1365}
\nonumber\\
&&\qquad\qquad\qquad
-\frac{1190473(n_1n_2)^3(n_1\!+\!n_2)^3}{455}
+\frac{19486(n_1n_2)^3(n_1\!+\!n_2)^2}{7}
\nonumber\\
&&\qquad\qquad\qquad
-\frac{1321279(n_1n_2)^3(n_1\!+\!n_2)}{1365}
+\frac{1082842(n_1n_2)^2(n_1\!+\!n_2)^4}{1365}
\nonumber\\
&&\qquad\qquad\qquad
-\frac{321112(n_1n_2)^2(n_1\!+\!n_2)^3}{195}
+\frac{372679(n_1n_2)^2(n_1\!+\!n_2)^2}{273}
\nonumber\\
&&\qquad\qquad\qquad
-\frac{82664(n_1n_2)^2(n_1\!+\!n_2)}{195}
-\frac{5528(n_1n_2)(n_1\!+\!n_2)^5}{91}
\nonumber\\
&&\qquad\qquad\qquad
+\frac{22112(n_1n_2)(n_1\!+\!n_2)^4}{91}
-\frac{175514(n_1n_2)(n_1\!+\!n_2)^3}{455}
\nonumber\\
&&\qquad\qquad\qquad
+\frac{384196(n_1n_2)(n_1\!+\!n_2)^2}{1365}
-\frac{21421(n_1n_2)(n_1\!+\!n_2)}{273}\; .
\label{eq:poly-2}
\end{eqnarray}
Two simple properties are satisfied by all these polynomials~:
\begin{equation}
{\cal P}_{2l}(n_1,0)={\cal P}_{2l}(0,n_2)=0\quad,\qquad {\cal P}_{2l}(1,1) = \frac{1}{3}\; .
\end{equation}
The first one is a consequence of the fact that if $n_1$ or $n_2$ is
zero, then all the closed paths one can construct have a vanishing
area. The second one follows from the fact that for $n_1=n_2=1$, all
the closed paths have area $-1$, $0$ or $+1$, and therefore contribute
equally to all the even moments.

The practical limitations of this algorithm are the growth of the number
of terms that need to be summed, and the dramatic cancellations that
occur among these terms: the intermediate calculations contain
polynomials with rational coefficients whose representation involves
very large integers, while the final result has a rather moderate
complexity. This suggests that there may be a better way to organize
the calculation, that would help avoiding these cancellations.

\section{Conclusions}
In this paper, we have used the worldline representation of scalar QED
lattice propagators in two dimensions in order to obtain the general
structure of the moments of the distribution of the areas of closed
random walks that make fixed number of steps in the two directions.
We find that these moments are the product of the number of such random
walks, times a polynomial in the number of steps made in each
direction. The derivation of this formula also provides an algorithm
to obtain this polynomial explicitly (although this is practical only
for moments of low order).

In this approach, one must choose a ``gauge'' to represent the
transverse magnetic field whose flux measures the areas, and there are
infinitely many ways of doing this. Although the final result --~in
particular the polynomials ${\cal P}_{2l}$~-- is gauge invariant, each
gauge choice may lead to an alternative to eqs.~(\ref{eq:pol1}),
(\ref{eq:pol2}) and (\ref{eq:pol3}) for representing this polynomial,
thereby potentially providing  a more efficient way of computing it.

\section*{Acknowledgements}
We would like to thank M.~Bauer and P.~Di~Francesco for useful
comments.  This work is supported by the Agence Nationale de la
Recherche project 11-BS04-015-01.

\appendix

\section{Combinatorial identities}
\label{app:combi}
\subsection{Derivation of eq.~(\ref{eq:comb1})}
\label{app:comb1}
Let us consider the following function,
\begin{equation}
F(x)\equiv \sum_{B\ge 0} x^B \sum_{b_0+\cdots+b_k=B}\frac{(a_0+b_0)!}{b_0!}\cdots\frac{(a_k+b_k)!}{b_k!}\; .
\end{equation}
By construction, the left hand side of eq.~(\ref{eq:comb1}) is the
$B$-th Taylor coefficient of $F(x)$. The sum on $B$ unconstrains the
sums over $b_0,b_1\cdots b_k$, so that we can write
\begin{equation}
F(x)=
\prod_{i=0}^k \left(\sum_{b\ge 0}x^b\frac{(a_i+b)!}{b!}\right)
=\prod_{i=0}^k\frac{a_i!}{(1-x)^{1+a_i}}
=\frac{a_0!\cdots a_k!}{(1-x)^{1+k+A}}\; ,
\end{equation}
where $A\equiv a_0+\cdots+a_k$. The Taylor coefficient of order $B$
can then be obtained as $F^{(B)}(0)/B!$, which gives immediately the
right hand side of eq.~(\ref{eq:comb1}).

\subsection{Derivation of eq.~(\ref{eq:comb2})}
\label{app:comb2}
Consider the function
\begin{equation}
G(x,y)\equiv\sum_{n_1,n_1'\ge 0}x^{n_1}y^{n_1'}
\sum_{\ontop{\alpha_0+\cdots+\alpha_k=n_1}{\beta_0+\cdots+\beta_k=n_1'}}
\!\!\!\!\!
\frac{(\alpha_0\!+\!\beta_0)!}{\alpha_0!\beta_0!}\cdots
\frac{(\alpha_k\!+\!\beta_k)!}{\alpha_k!\beta_k!}\; .
\end{equation}
The left hand side of eq.~(\ref{eq:comb2}) is the coefficient of
$(xy)^{n_1}$ in the Taylor expansion of $G(x,y)$. This function
can be rewritten as
\begin{eqnarray}
G(x,y)=\left(\sum_{\alpha,\beta\ge0}x^\alpha y^\beta\;\frac{(\alpha+\beta)!}{\alpha!\beta!}\right)^{1+k}=\frac{1}{(1-x-y)^{1+k}}\; .
\end{eqnarray}
The right hand side of eq.~(\ref{eq:comb2}) is then obtained as $\big[\partial_x^{n_1}\partial_y^{n_1}G(x,y))/n_1!^2\big]_{x,y=0}$.

\subsection{Derivation of eq.~(\ref{eq:comb3})}
\label{app:comb3}
In eq.~(\ref{eq:comb3}), we need to perform the sum over the $3^k$
possible assignments for the indices
$\epsilon_1,\cdots,\epsilon_k$. However, the summand depends on these
indices only via the numbers $n_-,n_0,n_+$ defined in eq.~(\ref{eq:nnn}).
Therefore, the sum over the $\epsilon_i$'s can be rewritten as follows
\begin{equation}
\sum_{(\epsilon_1,\cdots,\epsilon_k)\in\{-1,0,+1\}^k}\big(\cdots\big)
=
\sum_{n_-+n_0+n_+=k}\frac{k!}{n_-!n_0!n_+!}\;\big(\cdots\big)\; .
\end{equation}
(In the right hand side, the combinatorial factor counts the number
of assignments of the $\epsilon_i$'s that lead to a given triplet
$(n_-,n_0,n_+)$.) Therefore, the sum that appears in the right hand
side of eq.~(\ref{eq:comb3}) is equal to
\begin{eqnarray}
&&
\sum_{(\epsilon_1,\cdots,\epsilon_k)\in\{-1,0,+1\}^k}\!\!\!\!
(-1)^{n_0}
\frac{(2n_2\!-\!n_+\!-\!n_-)!}
{\big(n_2\!-\!n_+\big)!\big(n_2\!-\!n_-\big)!}=\nonumber\\
&&\quad=
k!\sum_{n_0=0}^k\!\frac{(-1)^{n_0}}{n_0!}\frac{(2n_2+n_0-k)!}{n_2!^2}
\!\!\!\!\!\!
\underbrace{
\sum_{n_-+n_+=k-n_0}\!\!
\frac{n_2!}{n_+!(n_2-n_+)!}\frac{n_2!}{n_-!(n_2-n_-)!}
}_{\frac{(2n_2)!}{(k-n_0)!(2n_2+n_0-k)!}}
\nonumber\\
&&\quad=
\frac{(2n_2)!}{n_2!^2}\sum_{n_0=0}^k(-1)^{n_0}\;\frac{k!}{n_0!(k-n_0)!}\propto (1-1)^k=0\quad\mbox{if }k\ge 1\; .
\end{eqnarray}

\subsection{Derivation of eq.~(\ref{eq:comb4})}
\label{app:comb4}
Consider the function
\begin{equation}
H(x,y)\equiv\sum_{n_1,n_1'\ge 0}x^{n_1}y^{n_1'}
\sum_{\ontop{\alpha_0+\cdots+\alpha_k=n_1}{\beta_0+\cdots+\beta_k=n_1'}}
\!\!\!\!\!
\frac{(\alpha_0\!+\!\beta_0)!}{\alpha_0!\beta_0!}\cdots
\frac{(\alpha_k\!+\!\beta_k)!}{\alpha_k!\beta_k!}\; (\alpha_k-\beta_k)^2\;.
\end{equation}
The left hand side of eq.~(\ref{eq:comb4}) is the coefficient of
$(xy)^{n_1}$ in the Taylor expansion of $H(x,y)$. Using some
results of the subsection \ref{app:comb2}, we first obtain
\begin{equation}
H(x,y)=\frac{1}{(1-x-y)^k}\;
\sum_{\alpha,\beta\ge 0}\frac{(\alpha+\beta)!}{\alpha!\beta!}(\alpha-\beta)^2\;x^\alpha y^\beta\; .
\end{equation}
The remaining sum in the right hand side of the previous equation is
given by
\begin{eqnarray}
\sum_{\alpha,\beta\ge 0}\frac{(\alpha+\beta)!}{\alpha!\beta!}(\alpha-\beta)^2\;x^\alpha y^\beta
&=&
(x\partial_x-y\partial_y)^2
\sum_{\alpha,\beta\ge 0}\frac{(\alpha+\beta)!}{\alpha!\beta!}\;x^\alpha y^\beta
\nonumber\\
&=&\frac{x^2+y^2-6xy+x+y}{(1-x-y)^3}\; .
\end{eqnarray}
This leads to the following expression for $H(x,y)$,
\begin{equation}
H(x,y)=\frac{1}{(1-x-y)^{1+k}}-\frac{3}{(1-x-y)^{2+k}}+\frac{2(1-4xy)}{(1-x-y)^{3+k}}\; .
\label{eq:comb4-1}
\end{equation}
From this formula, it is easy to extract the Taylor coefficient that
gives the sum of eq.~(\ref{eq:comb4})~:
\begin{eqnarray}
&&\frac{(2n_1+k)!}{n_1!^2\,k!}
-\frac{3(2n_1+k+1)!}{n_1!^2\,(k+1)!}
+\frac{2(2n_1+k+2)!}{n_1!^2\,(k+2)!}
-\frac{8(2n_1+k)!}{(n_1-1)!^2\,(k+2)!}=
\nonumber\\
&&\qquad=\frac{2kn_1}{(k+2)!}\frac{(2n_1+k)!}{n_1!^2}\; .
\end{eqnarray}

\subsection{Derivation of eq.~(\ref{eq:comb5})}
\label{app:comb5}
This time, consider the function
\begin{equation}
I(x,y)\equiv\sum_{n_1,n_1'\ge 0}x^{n_1}y^{n_1'}
\sum_{\ontop{\alpha_0+\cdots+\alpha_k=n_1}{\beta_0+\cdots+\beta_k=n_1'}}
\!\!\!\!\!
\frac{(\alpha_0\!+\!\beta_0)!}{\alpha_0!\beta_0!}\cdots
\frac{(\alpha_k\!+\!\beta_k)!}{\alpha_k!\beta_k!}\; (\alpha_i-\beta_i)(\alpha_j-\beta_j)\;.
\end{equation}
Following the same reasoning as in the previous appendix, we have
\begin{eqnarray}
I(x,y)&=&\frac{1}{(1-x-y)^{k-1}}\left[(x\partial_x-y\partial_y)\frac{1}{1-x-y}\right]^2
\nonumber\\
&=& \frac{(x-y)^2}{(1-x-y)^{3+k}}\nonumber\\
&=&
\frac{1}{(1-x-y)^{1+k}}-\frac{2}{(1-x-y)^{2+k}}+\frac{1-4xy}{(1-x-y)^{3+k}}\; .
\end{eqnarray}
The coefficient of $(xy)^{n_1}$ in the Taylor expansion of this function is
\begin{eqnarray}
&&\frac{(2n_1+k)!}{n_1!^2\,k!}
-\frac{2(2n_1+k+1)!}{n_1!^2\,(k+1)!}
+\frac{(2n_1+k+2)!}{n_1!^2\,(k+2)!}
-\frac{4(2n_1+k)!}{(n_1-1)!^2\,(k+2)!}=
\nonumber\\
&&\qquad=-\frac{2n_1}{(k+2)!}\frac{(2n_1+k)!}{n_1!^2}\; .
\end{eqnarray}

\section{Link with the Hofstadter-Harper Hamiltonian}
\label{app:HHH}
The purpose of this appendix is to establish the ``dictionnary''
between our lattice QED approach and the Hofstadter-Harper
Hamiltonian. Let us start from eq.~(\ref{eq:invprop-landau}), that
defines the inverse propagator in the Landau gauge. In this gauge, the
link variables that describe the background field on the lattice do
not depend on the coordinate along the direction $2$, and it is
therefore convenient to perform a Fourier transform in this variable
by writing
\begin{eqnarray}
f_{ij}&\equiv&\int_0^{2\pi}\frac{d\nu}{2\pi}\;e^{i\nu j}\; \tilde{f}_{i\nu}
\nonumber\\
D_{ij,kl}&\equiv&\int_0^{2\pi}\frac{d\nu d\nu'}{(2\pi)^2}\;e^{i(\nu j-\nu'l)}\;
  \widetilde{D}_{i\nu,k\nu'}\; .
\end{eqnarray}
From eq.~(\ref{eq:invprop-landau}), it is immediate to obtain an
explicit form for the Fourier transform of the inverse propagator~:
\begin{equation}
  {\bf a}^{2}\;\widetilde{D}_{i\nu,k\nu'}
  =
  2\pi\delta(\nu-\nu')\;\Big[4\delta_{i,k}-
    \big(\underbrace{h_1(\delta_{i,k+1}+\delta_{i,k-1})+2h_2\cos(\phi i+\nu)\delta_{i,k}}_{H^{(\phi,\nu,h_1,h_2)}_{ik}}\big)\Big]\; ,
\end{equation}
where the underlined operator $H^{(\phi,\nu,h_1,h_2)}$ is the
(anisotropic) Hofstadter-Harper Hamiltonian \cite{Harpe1,Hofst1} for a
magnetic flux $\phi$ and a wavenumber $\nu$ in the direction $2$. The
proportionality to $\delta(\nu-\nu')$ is specific to Landau gauge, and
ensures that the inverse can be calculated separately in each $\nu$
sector. The Fourier transform $\widetilde{G}$ of the propagator is
defined by
\begin{equation}
  \sum_{k\in\mathbbm{Z}}\int_0^{2\pi}\frac{d\nu'}{2\pi}
  \;\widetilde{D}_{i\nu,k\nu'}\widetilde{G}_{k\nu',l\nu''}
  =2\pi\delta(\nu-\nu'')\;\delta_{i,l}\; ,
\end{equation}
and it can be formally written as
\begin{equation}
  \widetilde{G}_{i\nu,k\nu'}=
  \frac{{\bf a}^2}{4}\,2\pi\delta(\nu-\nu')\,\sum_{n=0}^\infty\left(\frac{H^{(\phi,\nu,h_1,h_2)}}{4}\right)^n_{ik}\; .
\end{equation}
From this expression, the diagonal elements of the propagator in
direct space are readily obtained, and by comparing with
eq.~(\ref{eq:G00}) we get
\begin{equation}
\sum_{n_1,n_2=0}^\infty
\left(\frac{h_1}{4}\right)^{2n_1}
\left(\frac{h_2}{4}\right)^{2n_2}
\sum_{\gamma\in{\bs\Gamma}_{n_1,n_2}}e^{i\phi\,{\rm Area}\,(\gamma)}
=
\sum_{n=0}^\infty\int_0^{2\pi}\frac{d\nu}{2\pi}\;\left(\frac{H^{(\phi,\nu,h_1,h_2)}}{4}\right)^n_{ii}\; .
\label{eq:HH-ident}
\end{equation}
Note that any value of the coordinate $i$ may be used in the right
hand side of this equation, since the problem is invariant under
discrete translations on the lattice.

Since the Hamiltonian $H^{(\phi,\nu)}$ is the sum of two terms
respectively proportional to $h_1$ and $h_2$, eq.~(\ref{eq:HH-ident})
leads to the following identity
\begin{equation}
\sum_{\gamma\in{\bs\Gamma}_{n_1,n_2}}e^{i\phi\,{\rm Area}\,(\gamma)}
=
\int_0^{2\pi}\frac{d\nu}{2\pi}\;\left({H_1+H_2^{(\phi,\nu)}}\right)^{2n_1,2n_2}_{ii}\; ,
\label{eq:exp-moment}
\end{equation}
where
\begin{eqnarray}
  \left(H_1\right)_{ik}\equiv\delta_{i,k+1}+\delta_{i,k-1}
  \quad,\qquad
  \left(H_2^{(\phi,\nu)}\right)_{ik}\equiv 2\cos(\phi i+\nu)\,\delta_{i,k}
\end{eqnarray}
and where the notation $(H_1+H_2^{(\phi,\nu)})^{2n_1,2n_2}$ denotes
the sum of all the terms with $2n_1$ powers of $H_1$ and $2n_2$ powers
of $H_2^{(\phi,\nu)}$ in the $2(n_1+n_2)$-th power of
$H_1+H_2^{(\phi,\nu)}$. The main difficulty in calculating the right
hand side is that $H_1$ and $H_2^{(\phi,\nu)}$ do not commute (except
in the trivial case $\phi=0$), and each term in this sum depends on
how the $H_2^{(\phi,\nu)}$'s are interspersed between the
$H_1$'s. The expansion rules described in the subsection
\ref{sec:rules} provide the necessary bookkeeping for performing this
calculation. In particular, the quantities $\sigma_i\equiv
\sum_{j=1}^i\epsilon_j$ defined in eq.~(\ref{eq:sigma}) record the
cummulative effect of the $\phi$'s up to the $i$-th factor $H_1$.

The main result of this paper, namely the eq.~(\ref{eq:moments-0}),
does not apply directly to the left hand side of
eq.~(\ref{eq:exp-moment}) but to its expansion in powers of the flux
$\phi$. Therefore, it provides only indirect information on the
structure of the trace of the powers of the Hofstadter-Harper
Hamiltonian: if one keeps track separately of the number of powers
$n_1$ and $n_2$ of $H_1$ and $H_2^{(\phi,\nu)}$ respectively, then the
Taylor coefficients of the expansion in $\phi$ are a combinatorial
factor times a polynomial. The summation over all $n_1+n_2=n$ then
leads to the structure established in ref.~\cite{MingoN1}, namely
$(2n)!^2/n!^4$ times a rational fraction in $n$. Using a computer
algebra system such as Maple and the expressions listed in
eqs.~(\ref{eq:poly}), (\ref{eq:poly-1}) and (\ref{eq:poly-2}), one can
obtain the explicit form of this rational fraction for the lowest
order moments, but these expressions are not particularly
illuminating.

Because it only applies to the expansion of eq.~(\ref{eq:exp-moment})
in powers of $\phi$, it also seems difficult to connect our result to
other known results about the spectrum of the Hofstadter-Harper
Hamiltonian and in particular to the distinctive differences that
arise depending on whether the flux is commensurate with $2\pi$ or
not. Indeed, the periodicity of the Hamiltonian in the coordinate $i$
when the flux is $2\pi$ times a rational number is a ``non
perturbative'' property which is only manifest if one does not expand
the cosine in powers of $\phi$. Going beyond our result for the
moments would require to perform the sum on the moment order $2l$ in
eq.~(\ref{eq:moments-0}) (which does not seem feasible given the fact
that the polynomial ${\cal P}_{2l}$ is not known explicitly for all
$l$), or to perform the sum over the $\alpha_i$'s and $\beta_i$'s in
eq.~(\ref{eq:pert-2}) without first expanding the exponential that
contains the flux.


\end{document}